# Pay-with-a-Selfie, a human-centred digital payment system

Ernesto Damiani (*), Perpetus Jacques Houngbo (^), Rasool Asal (*), Stelvio Cimato (°), Fulvio Frati (°), Joel T. Honsou (^) , Dina Shehada (*), Chan Yeob Yeun (*)

*(\*) EBTIC-Khalifa University, Abu Dhabi, UAE*
  *Abu Dhabi Campus PO Box 127788 Abu Dhabi - UAE*
  *Fax: +44 (0)1908 861 157*
  *{ernesto.damiani, rasool.asal, dina.shehada, cyeun}@kustar.ac.ae*

*(°) Dipartimento di Informatica, Università degli Studi di Milano, Italy*
  *via Bramante, 65 26013 Crema (CR) – Italy*
  *Fax: +39 0373 898010*
  *{stelvio.cimato, fulvio.frati}@unimi.it*

*(^) Institut de Mathematique et Science Physique*
  *Quartier Avakpa, BP 613, Porto-Novo, Bénin*
  *Fax : +229 20 22 24 55*
  *joelhoun@gmail.com, jacques.houngbo@auriane-etudes.com*

**Corresponding author: Ernesto Damiani, ernesto.damiani@kustar.ac.ae**

Provide short biographical notes on all contributors here if the journal requires them.

# Pay-with-a-Selfie, a human-centred digital payment system

*Mobile payment systems are increasingly used to simplify the way in which money transfers and transactions can be performed. We argue that, to achieve their full potential as economic boosters in developing countries, mobile payment systems need to rely on new metaphors suitable for the business models, lifestyle, and technology availability conditions of the targeted communities.*

*The Pay-with-a-Group-Selfie (PGS) project, funded by the Melinda & Bill Gates Foundation, has developed a micro-payment system that supports everyday small transactions by extending the reach of, rather than substituting, existing payment frameworks. PGS is based on a simple gesture and a readily understandable metaphor. The gesture – taking a selfie – has become part of the lifestyle of mobile phone users worldwide, including non-technology-savvy ones. The metaphor likens computing two visual shares of the selfie to ripping a banknote in two, a technique used for decades for delayed payment in cash-only markets. PGS is designed to work with devices with limited computational power and when connectivity is patchy or not always available. Thanks to visual cryptography techniques PGS uses for computing the shares, the original selfie can be recomposed simply by stacking the shares, preserving the analogy with re-joining the two parts of the banknote.*

## 1. Introduction

In the vision of the "mobile wallet revolution", consumers have a smartphone in the pocket instead of a regular wallet, and use their own device to make payments or transfer money to each other. In the last fifteen years, many apps (such as Google Wallet, or Samsung Pay) and technologies (like NFC) have become available supporting smartphone users in their payment needs (Brazier 2013).

In developing economies, the number of people with access to mobile phones is high and increasing from year to year, but the availability of the overall technological infrastructures and the usage habits are very different from those of the western world.

A recent report[1] shows that the diffusion of cell phones in Nigeria and South Africa is equal to the one of the United States, where about 90 percent of adults owns a mobile phone. A big share of those cell phones, however, are low-end models, with capabilities of calling, texting and basic Internet browsing. Also, voice coverage does not automatically imply Internet access. Cellular data network access is location-wise patchy, because cell data coverage is still being completed in many countries. The GSMA[2] expects the penetration rate of mobile Internet in Sub Saharan Africa to reach a level (40%) by 2020. That is still relatively low in absolute terms. Perhaps more importantly, individuals perceive connections as *time-wise* patchy, as SIMs are often shared among members of the same family or social group.

Nevertheless, easy access to voice and textual communication is having a deep impact in the local economies. Google's Project Loon[3], whereby helium balloons are deployed and float to an altitude of a few kilometres can extend GSM coverage and even offer 4G connections in many remote locations, some yet to be reached by traditional utilities like electric power and water.

The availability of financial, health-care, agricultural, and educational services via mobile terminals is changing traditional relationships inside previously remote local communities, and it is bringing increasing economic opportunities. For example, many people living in rural areas of Africa and Asia have started using SMS services to find out daily prices of agricultural goods, to improve their bargaining position in local markets, and to select markets that offer the maximum income (Kochi 2012).

---

[1] Cell Phones in Africa: Communication Lifeline, Pew Research Center, https://goo.gl/Y7sWwg

[2] The GSMA represents the interests of mobile operators worldwide, uniting nearly 800 operators with more than 250 companies in the broader mobile ecosystem, including handset and device makers, software companies, equipment providers and Internet companies, as well as organisations in adjacent industry sectors. http://www.gsma.com/

[3] https://www.solveforx.com/loon/

A number of mobile digital-money platforms have become available, supporting small money transfers among users. Airtime and MPESA (Vyas, Gau and Singh 2016) are two examples of mobile cash transfer systems used in several countries; additional systems will be reviewed in Section 5.

PGS design starts from the notion that the impact of traditional payment systems' in developing countries is impaired by two hidden assumptions that – though reasonable in the western world – may prove to be partially flawed for emerging economies. The first assumption is "always-on" usage of cellular data connections. We argue that even when cellular data network coverage will be completed, SIM sharing and other spontaneous cost reduction practices will cause emerging countries' users to be online only occasionally during the day.

Our approach envisions a *human-centered* design (Khosla, Damiani and Grosky 2003) that accompanies (and takes advantage of) *spontaneously emerging* usage patterns rather than trying to oppose them. The second assumption we consider debatable is the *cultural ubiquity* of the credit-card metaphor for deferred payments. Credit cards were born as a way for affluent diners to defer payments of restaurant meals. According to some[4], the credit card underlying metaphor was conceived in 1949 when a man named Frank McNamara had an expensive business dinner in the fashionable Major's Cabin Grill restaurant in New York. When the bill arrived, McNamara realized he had forgotten his wallet, and wondered if there could be an alternative to paying cash. McNamara and his partner, Ralph Schneider, returned to

---

[4] See for instance Encyclopedia Britannica at https://www.britannica.com/topic/credit-card

Major's Cabin Grill in February of 1950 and paid the check with a small, cardboard card. The story conveys clearly the original metaphor underlying the invention: the buyer shows to the seller the equivalent of a gentleman's visit card (endorsed by an affluent third party) as a proof of his/her solvency and willingness to pay later. The visit card-based metaphor looks hardly a good fit for the culture of cash-centred marketplaces in emerging countries, where written documents are not always trustworthy and the link between a card and the identity of the cardholder may be perceived as weak (Damiani, De Capitani di Vimercati and Samarati 2003).

This paper presents a radically alternative approach, Pay-with-a-Group Selfie (PGS). PGS is a mobile payment technique that leverages on face-to-face exchanges where pictures taken with mobile phones are used to embed all information (the parties' identities, the exchanged goods/services, and the price) of a business transaction. The system relies on Visual Cryptography (VC) to generate two untamperable shares of the selfie showing the transaction. The buyer and the seller hold a share each, and the transaction can be checked simply by stacking the shares.

The paper is organized as follows: Section 2 briefly reviews the foundations of visual cryptography. Our aim is to make this paper self-contained by providing the non-specialist reader with the notions needed to understand PGS operation; a complete survey of applications of VC can be found in (Cimato and Yang 2011). Section 3 describes PGS design principles and architecture. Section 4 discusses the current prototype design while Section 5 deals with interfacing PGS with existing mobile money systems. Section 6 discusses overloading a traditional role of the informal economy, the "ambulant banker" as part of the PGS deployment in the field, in the African Republic of Benin. Section 7 discusses the design on the experimental PGS deployment in Benin. Finally, Section 8 draws the conclusion and outlines our future work.

2.  **Visual cryptography**

Visual cryptography schemes allow the encoding of a secret image, usually consisting of black & white pixels, into $n$ shares that are distributed to a set of $n$ participants. The secret pixels are shared relying on techniques that intelligently subdivide each secret pixel into a certain number of subpixels. Each share is then composed of black and white subpixels, which are printed in close proximity to each other, so that the human visual system averages their individual black/white contributions and make it possible the reconstruction of the secret image. White colour is associated to a transparent pixel, so that the superposition of white pixels lets the colour of the pixel contained in the other shares pass through.

The shares are such that only qualified subsets of participants can "visually" recover the secret image, but other subsets of participants cannot gain any information about the secret image by examining their shares. The shares can be conveniently represented with an $n$ x $m$ matrix $S$, which is called *distribution matrix*; each row of this matrix represents one share, i.e., $m$ subpixels, and each element is either *0*, for a white subpixel, or *1* for a black subpixel. To reconstruct the secret image, participants need only to stack together their shares. The grey level of the combined share, obtained by stacking the transparencies, is proportional to the *Hamming weight w(V)* of the *m*-vector $V = OR(r_{i1} ; \ldots ; r_{is})$, where $r_{i1} ; \ldots ; r_{is}$ are the rows of $S$ associated with the stacked transparencies. This grey level is interpreted by the human visual system as black or as white in according with some rules of contrast. Since each secret pixel is represented by $m$ pixels in the shares, the reconstructed image will be bigger than the original.

Visual cryptography schemes are characterized by two parameters: *pixel expansion*, corresponding to the number of subpixels contained in each share (transparency) and *contrast*, which measures the "difference" between a black and a white pixel in the reconstructed image. Other parameters include the *number of participants n*, the *threshold k* that determines whether a set of participants is qualified to reconstruct the image, *pixel expansion m,* and *contrast threshold*, which determines whether a reconstructed pixel is considered white or black. Naor and Shamir (Naor and Shamir 1994) introduced this cryptographic paradigm. They analysed the case of *(k,n)*-threshold visual cryptography schemes, in which a black and white secret image is visible if and only if at least k transparencies among n are stacked together. The original model by Naor and Shamir has been extended in (Ateniese, et al. 1996) to a general access structure, that explicitly specifies all qualified and forbidden subsets of participants; a number of variants have also been considered in literature, extending the scheme to process colour images (Cimato, De Prisco and De Santis 2007) or changing the reconstruction phase by introducing a probabilistic factor (Cimato, De Prisco and De Santis 2006).

To explain how a visual cryptographic scheme allows the generation of shares, a basic operation in PGS, we report in Figure 1 the original (2, 2) encoding scheme introduced by Naor and Shamir, where both 2 shares of the 2 participants are needed to reconstruct the secret image. The scheme operates by encoding each single pixel $p$ of a binary image into a corresponding pair of pixels included in the two shares $S_1$ and $S_2$. If $p$ is white, the dealer randomly can choose one of the first two rows of the table in Fig. 1 to build $S_1$ and $S_2$. If $p$ is black, the dealer randomly chooses one of the last two rows of the table. The probabilities of the two encodings are the same, independently of the original pixel being black or white. Thus, an adversary looking at a single share has no information about the original value of *p*. When the two shares are stacked together, if *p* is black, two black sub-pixels will appear, while, if *p* is white, one black sub-pixel and one white sub-pixel will appear as reported in the rightmost column of the table. The human visual system will distinguish whether *p* is black or white due to the contrast between the two reconstructed pixels, even if instead of a really white color, the color of the merged subpixel will be gray.

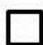

Figure 1: Naor and Shamir (2,2) VC scheme.

## 3. PGS Design

PGS is based on three design principles:

- Using a simple digital object (a selfie) for embedding all information about a purchase: the actors (buyer and seller), the product or service being sold and the price agreed upon between the parties.
- Producing secure VC shares of the original selfie, so that each single share reveals no information on the original selfie, but their combination using the human visual system reconstructs the original image and provide proof of the original purchase.
- Integrating purchase validation with any existing payment systems, as well as innovative systems exploiting virtual currencies. PGS is not intended to replace existing payment systems, but to decrease their cost-per-purchase and extend their reach where data connection is not continuous due to patchy network coverage, cost concerns or usage habits.

The idea of using a selfie showing all parties hints at the traditional way in which business is conducted in open-air markets. The human visual system has been used for countless years to establish the context of each sale: the purchaser, the supplier, the purchased goods or services, the price to be paid, as well as the time, and location of the transaction. We argue that non-technology-savvy users will be comfortable with this setting, as taking a selfie requires virtually no training or technology awareness. Furthermore, our proof of purchase is *self-validating*, just like a banknote, and like a banknote ripped in two it is held jointly by the two parties until the money transfer can be finalized.

Behind the scenes, PGS uses VC (Section 2) to split the selfie, generating two random-looking shares. As discussed in Section 2, this technique provides unconditional security (Naor and Shamir 1994): owning a single share gives no possibility to reconstruct the original image, and tampering with it makes the reconstruction impossible. Also, this technique supports the metaphor of putting together the two parts of a ripped banknote. In Section 2 we explained how the reconstruction of the selfie can be performed without a computer, simply by stacking the shares printed on two transparencies. In PGS implementation, reconstruction is performed by a desktop application that stacks the shares provided by the parties to validate the transaction.

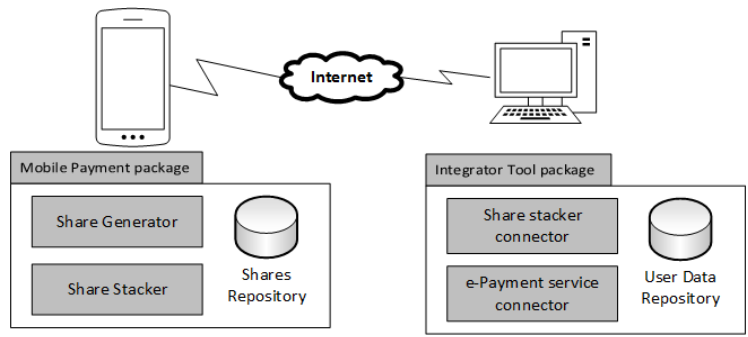

**Figure 2: Mobile payment framework component diagram.**

Finally, no "always on" assumption is made: when no network coverage is available, purchase validation is deferred to when shares will have been independently and securely delivered to a payment service provider (see Section 7), who will stack them and validate the purchase.

*3.1 PGS Architecture and Operation*

PGS architecture (Figure 2) includes three basic components: two mobile applications running on the seller and the buyer's smartphones and a desktop application running at the remote point of service.

The seller app takes the selfie, converts it to black and white, and generates the shares; then sends a share to the buyer app using a local Bluetooth connection.

The gray-scale representation of a selfie showing two parties concluding a transaction and the price is shown in Figure as well as the black and white version of the selfie. While VCS can support grayscale and even colour images, we fund that conversion to black and white makes shares' generation and transmission much faster, without impairing face recognition (Anisetti, et al. 2007).

Each PGS app is in charge of delivering its share to the remote desktop application, using an Internet connection once it becomes available or exploiting a human broker (Section 5). After both shares have arrived, our desktop application reconstructs the selfie and a human operator (or a software agent) validates the purchase and adds it to the queue toward a money transfer.

Once validated purchases have reached the desired threshold amount, the bank clerk uses the desktop application to interact with the payment service, ensuring that the supplier gets the money, and that the buyer gets the goods.

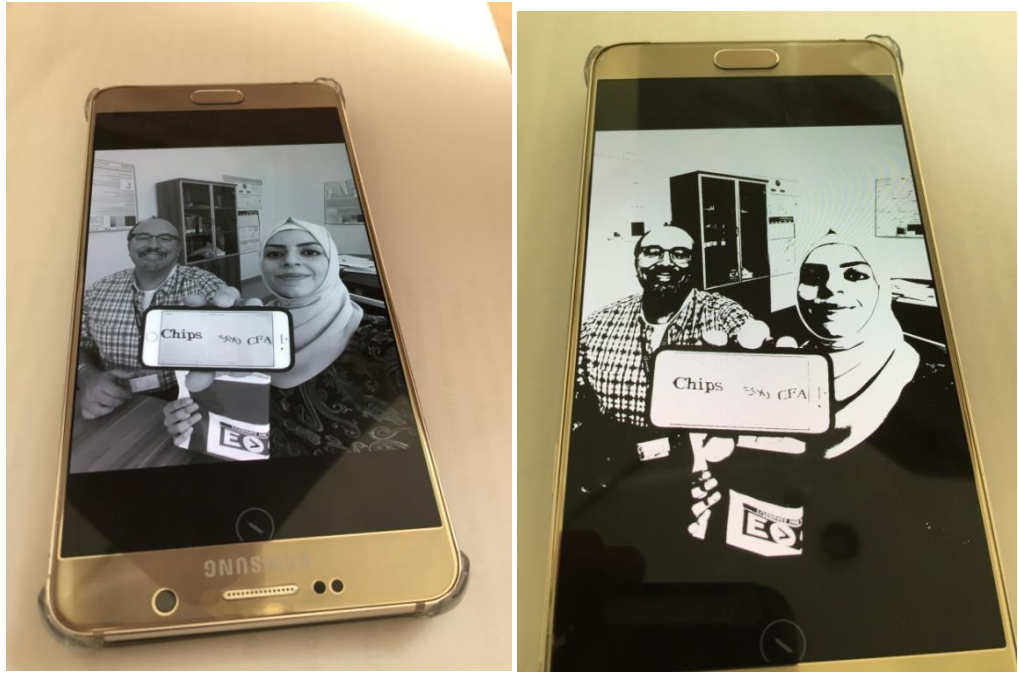

**Figure 3: The selfie showing the parties, and the price of the transaction. (a) shows the image in grayscale, (b) the same image converted in black and white.**

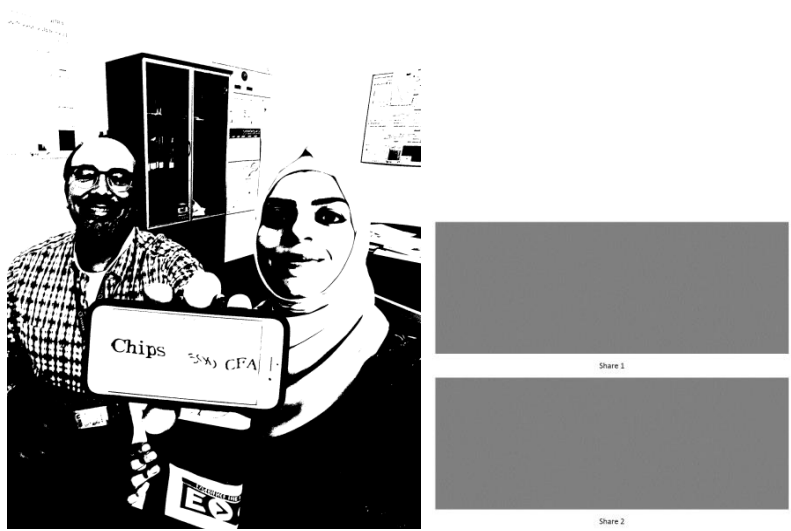

**Figure 4: The reconstruction phase. (a) The two shares generated by the app; (b) The reconstructed image.**

*3.2 PGS Business models*

The PGS system supports multiple ways to finalize a purchase, based on different business models. The choice of a business model may depend on the amount of money involved in the purchase as well as on the trustworthiness of the parties. The basic model is "*carry-then-cash*" (Caskey 1997): first, the selfie is taken, the goods (and the shares) are exchanged, and then the purchase is validated once the selfie's shares have been independently delivered to the point of service and validated. In this model, it is possible to keep payment transactions *de-coupled* from payments, allowing several validated purchases to accumulate before triggering a payment. This allows multiple transactions between the same parties to share the financial cost of an individual money transfer. Of course decoupling validation from payment requires a strong trust link between the buyer and the seller. While *purchases* are validated, *payments* are never guaranteed in PGS: even a validated purchase is not guaranteed to result in a successful payment, as the buyer may be declined by external payment system. The risk involved is acceptable for an individual purchase, as the seller is immediately informed of the refusal and the buyer is blacklisted. Accumulating validated purchases results in a lower financial cost per purchase, but also in a risk increase due to higher impact of insolvency. However, we claim that periodic deliveries of goods to the same buyer, e.g. a wholesaler baker supplying a bread reseller (see Section 5) may involve little insolvency risk and justify decoupling.

In the "*cash-then-carry*" model (Caskey 1997) taking the selfie and exchanging shares only express a *commitment* on the part of the seller to deliver the goods once the payment is done. The physical exchange of goods and money is finalized later at the point of service who – after validating the purchase – carries out the payment once the seller has made the goods available at its location. Payment being declined is not an issue here, as the goods are released once payment has gone through.

*3.3 Securing PGS Exchanges*

As we discussed in Section 2, VCS ensures untamperability and confidentiality of shares. We also remark that when taking the selfie in PGS the price is displayed using a *captcha* generator (von Ahn, et al. 2003) on the buyer's phone, in order to prevent the seller from doctoring the selfie using a malicious software *prior* to share generation. Without this precaution, a rogue seller could try and modify the price before the share generation process, e.g. using some automatic tool. Preventing automatic recognition of the price and setting a timeout for the share generation ensures that the shares are computed on the original selfie.

However, the overall security of the purchasing process is achieved in PGS more by social engineering than by technology. Immediate blacklisting on the part of the point of service is the main tool we envision to prevent all types of malicious behaviour on the part of the buyer and on the seller, such as not delivering a share to the point of service. Indeed the point-of-service (or the broker, see Section 5) acts as a trusted third party (the "village chief") who will discourage users' malicious behaviour in the long term rather than preventing it entirely. While the "village chief" is supposed to discourage rational attacks, we rely on local social stigma - as well as on the Supplier app having a special tamper-resistant installation and release - to prevent malicious behaviour.

**4.     PGS prototype**

PGS software development was carried out by three teams: one in charge of the client-side back end (generating the shares at the seller's device), one in charge of the pint-of-service back-end (reconstruction of the original selfie from the shares, validation of the purchase and interface with external payment systems), and one in charge of the user interface on the mobile devices.

Android was selected as the execution environment at the client side. According to Gartner[5] report for August 2016, as of Q2 2016, Android represents over 86% of market share and dominates the market together with iOS that counts for 13%. Also it is by far the preferred choice in Africa and in particular in the Republic of Benin, where PGS has been experimented.

**5. Interconnecting PGS with mobile money systems**

In this section we review choices for interfacing PGS with a payment system operating in the country of experimentation, Benin. We believe the situation in Benin to be typical of West Africa and more in general of a number of developing countries.

*5.1 State of the play of mobile money in Benin*

Digital financial services (DFS) market in Benin has not yet entered the rapid growth age that some were expecting. The central bank (BCEAO) enlists 15 banks in Benin. The main, the most visible or well-known of them are Bank of Africa, Ecobank, Diamond Bank, United Bank for Africa, and Société Générale. In the category of microfinance institutions, the list contains 54 entities. Also, Benin has five mobile network operators (MNOs) (Benin 2015). Only two of them, MTN and Moov, are currently offering mobile money services. Two other organization operate in mobile money: *(i)* the Post Office (Poste du Bénin) and *(ii)* an inter-market association, Association pour la Solidarité des Marchés du Bénin (ASMAB).

---

[5] http://www.gartner.com/newsroom/id/3415117

Table 1 Mobile network operators (MNOs) in Benin.

| Operators | Number of subscribers | Market share % |
|---|---|---|
| **MTN, SPACETEL BENIN SA** | 3 893 252 | 36 |
| **MOOV, ETISALAT-BENIN SA** | 3 586 006 | 33 |
| **GLO MOBILE BENIN SA** | 1 908 653 | 18 |
| **BELL BENIN COMMUNICATIONS** | 1 103 629 | 10 |
| **LIBERCOM** | 289 335 | 3 |
| **Total** | **10 780 875** | |

Let us now briefly review the available services for interconnection with PGS.

**Mobile Money/MTN**

MTN opened its mobile money services in September 2010 (ECOBANK 2013) with the financial partnership of Ecobank and the technical one of Fundamo and Gemalto. As of July 2015, Mobile Money claimed to have 4,000 agents and 900,000 subscribers (Sheree 2015). Services offered include money transfer from/to mobile phones, money deposit and withdrawal. Also, an *online account* enables the account holder to in make payments to anyone in Benin - no need for the recipient to have a MTN subscription or even a mobile phone. From the security point of view, MTN Mobile Money account has its own Personal Identification Number (PIN), which is different from the one of the corresponding SIM card, although the application is stored on the SIM.

**Flooz/Moov**

After launching the Flooz service on the mobile networks of its subsidiaries in Cote d'Ivoire, Niger and Togo, in December 2013 the global mobile operator Etisalat did the same for its Moov Benin subsidiary. Banque Atlantique du Bénin is the bank partner of Flooz (Barton 2006). Besides money deposit and withdrawal, services offered by Flooz include money transfer, purchase of airtime, payment of bills and purchase of goods and services with merchants who subscribed to the system.

Flooz subscription process consists in showing up in person at a Flooz agent location to fill up a form and provide photo ID.

**Mobile-based remittance at La Poste du Benin**

The share of financial services in the activities of La Poste du Benin is increasing year after year. La Poste du Benin has a remittance service, "Mandat flash", that can be seen as a model of simplicity. When the sender reaches an agent of La Poste du Benin, it only has to provide the name and the mobile phone number of the recipient, plus the amount to transfer. The receiver immediately gets an SMS with the information of the remittance alongside with a secret code to be used when collecting the money at the agent. The secret code is only necessary when the receiver cannot display an ID document.

**Mobile Money of Association pour la Solidarité des Marchés du Bénin (ASMAB)**

Association pour la Solidarité des Marchés du Bénin (ASMAB) is a microfinance institution with three main activities: *(i)* savings, *(ii)* loan reimbursement, and *(iii)* commitment by signature (endorsement, pledge, etc.). In terms of mobile money, ASMAB is implementing a magnetic card for secure savings CARMES (Carte Magnétique d'Epargne Sécurisée). CARMES intends to enable financial transactions by using mobile phones.

Two of the reviewed services (Mobile Money and Floov share a key feature: a bank handles mobile money transfers between users. Such banks are natural candidates for hosting the PGS point-of-service and converting purchases into money transfers. In turn, Mandat Flash and CARMES can support a looser integration, where the PGS point-of-service is run by a third party like a Non-Governmental Organization NGO, or by a market association.

## 6.     Role-driven deployment of PGS

As discussed in Section 3, the initial PGS deployment scheme relies on Internet connection eventually becoming available to perform file exchange after the selfie has been taken and the shares generated (Figure ). The shares transmission is handled via Android Update, the mechanism that Android devices use to find the latest updates for their operating system and get it.

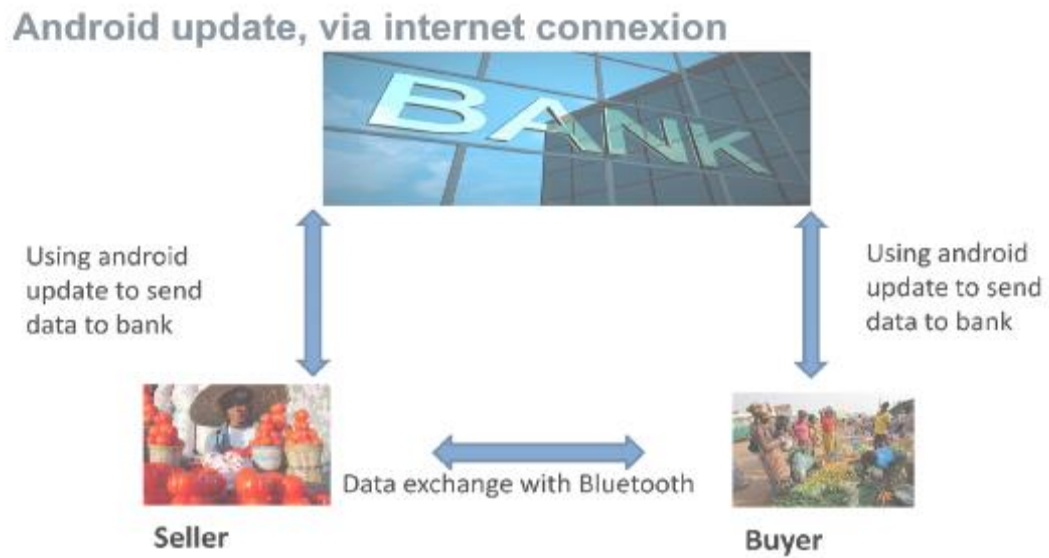

**Figure 5: Data exchange using android update, via internet connection**

Seminal work on handling user uncertainty in e-business environments (Pavlou, Liang and Xue 2006) has shown that relying on an intermittently available Internet connection to carry out a complex e-business transaction makes the participants uncertain and worried, possibly disrupting their perception of the underlying metaphors (Khosla, Damiani and Grosky 2003). In the case of PGS, both the seller and the buyer need to independently reach an area where Internet connection is available for transferring their shares to the bank. This introduces a time uncertainty. In order to alleviate it, in Benin we re-organized the whole system around the role of a *broker*. The broker-centred PGS system can then be represented as in Figure .

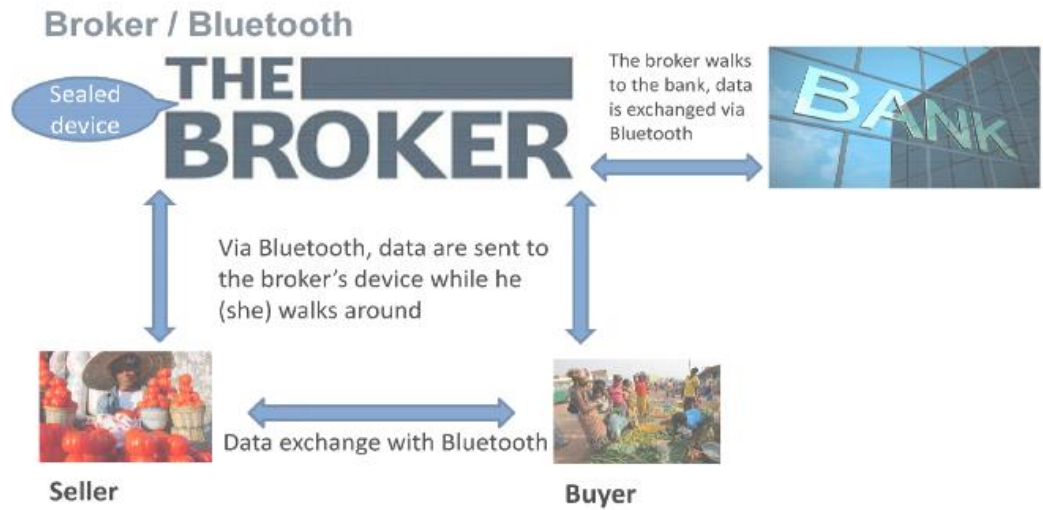

**Figure 6: Data exchange using Bluetooth and the services of a broker**

The role of broker is not an artificial one; rather it is a permanent part of the business ecosystem in many developing countries like Benin. This role is also known as "ambulant banker".

## 6.1 The ambulant banker

The role of "ambulant banker" (in French, "tontinier") is well established in all West African countries. Operating basically as deposit account (Kayodé 2006) for his customers, the "ambulant banker" collects everyday a fixed amount from his customer base. The business model is that the ambulant banker's payment is one day of collection over a period (*tontine*) of a month. After payment of the service to the ambulant banker, people can withdraw from him/her their money whenever needed. This is often the case for emergencies like illnesses, burial ceremonies of relatives. Since the ambulant banker is in possession of some amount of funds, he/she often uses them for money loans on a very short-term basis. Today, the ambulant banker is fully connected to the formal banks where he/she may hold accounts or make deposits (Mayoukou 1996).

In order to sustain his business, the ambulant banker tends to market the quality of his lending services, building a brand as a lender of last resort, a "financial saviour" for his customers. His business *motto* usually includes concepts like credibility, reliance, seriousness, and proximity (Mayoukou 1996). Figure shows the value chain that links the ambulant banker to his customers and cements their relation based on trust and reputation.

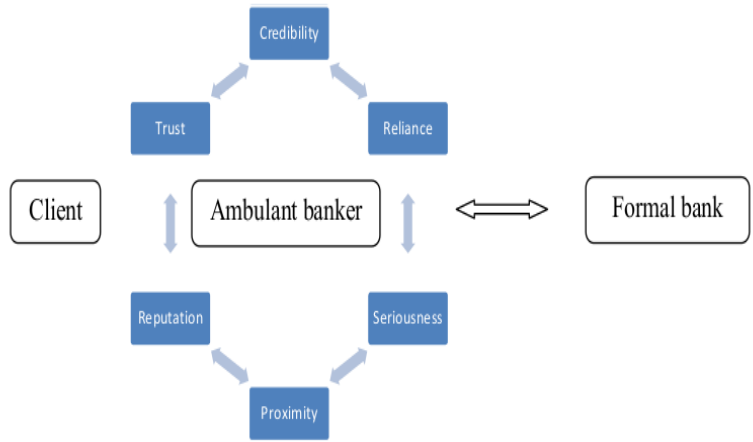

Figure 7: The trust network of an ambulant banker

## 6.2 Overloading the ambulant banker role

The ambulant banker operates as an intermediary in the informal financial system, smoothly connecting the informal economy to the formal one. As the ambulant banker already holds a network of trust for him/herself, we overloaded his role making the ambulant banker a pivot of the PGS system's human centred design. Specifically he/she will act as a human data storage device, transferring data between buyers, sellers, and their bank. In other words, the ambulant bank acts as a human *data broker* of the PGS system, operating between the seller/buyer pair and the bank as shown in Figure .

Like his ICT counterparts, the human data broker needs to be reliable and trustworthy. We take advantage of ambulant bankers to be seen as a safe for poor people, acting as a deposit bank and a lender of last resort, making him/her the most trusted role in the informal market economy. In terms of the PGS metaphor, the ambulant bankers is trusted to hold the two parties of the ripped banknote representing the payment for each transaction, and to put them together only at the bank, where the banknote is recomposed and deposited on the account of the seller. Instead of asking people to change their behaviour to operate digital devices, the PGS system accommodates their current usage patterns and roles, exploiting the trust network that is sustaining the informal economy.

## 7. The Bank Application

We now provide some details on PGS Bank application. It was developed to provide a simple and user-friendly interface to upload ongoing transactions, approve/reject completed transactions, and export the corresponding data in CSV and XLS format. In order to fulfil all those goals, the application has been developed as a web application.

The bank application provides different views according to the user logging into the system, which can be Admin or a normal User. Users can upload the shares, and display the state of the purchase, which is automatically processed by the system starting a background job.

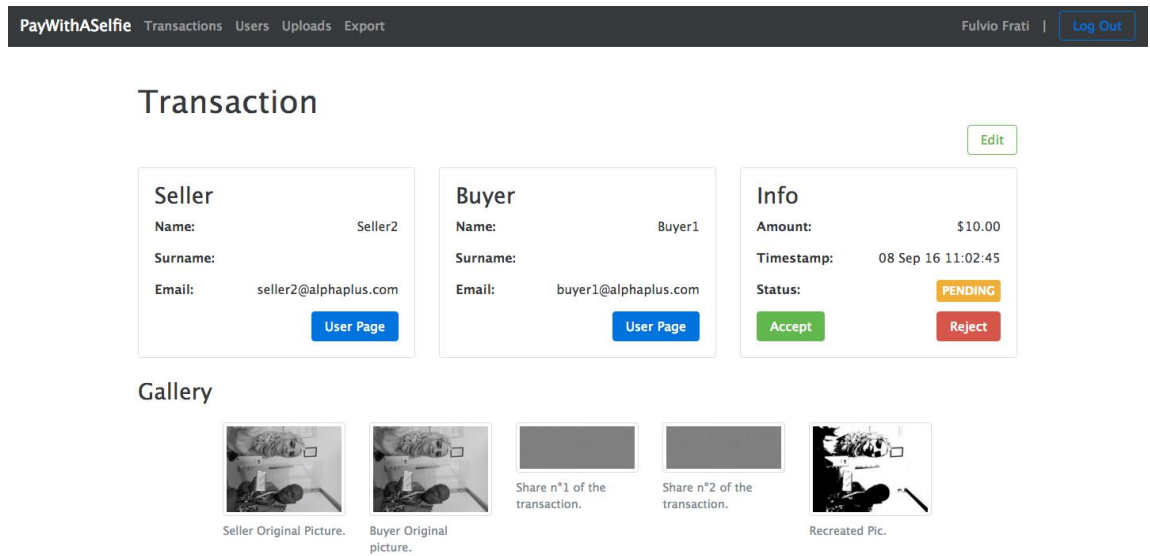

Figure 8: Bank application landing page.

Figure 9: Bank application: list of transactions.

The Admin user can access all the purchases in the system, listed according to their state (Accepted, Rejected, Incomplete) and validate each purchase, after checking all the details (the shares, the reconstructed images, the users, etc. ) that are displayed in a separate page. The Bank application relies on the Ruby-on-Rails framework[6] allowed releasing the Application Program Interface (APIs used by the PGS Android apps to upload the purchase shares. These APIs rely on the standard OAuth2 framework[7] to authenticate the apps.

## 8. Experimentation of PGS

The experimentation phase has defined the following objectives:

- assessment of the viability of PGS design;
- functional testing and experimentation;

---

[6] http://rubyonrails.org

[7] https://oauth.net

- performance evaluation;
- user acceptance;
- checking of unexpected results.

The experimentation spanned 15 days, covering two local markets in Cotonou (Benin) and involved ten sellers in order to reach a threshold of 1500 validated purchases.

The experimentation was supervised by the department of IT training at Institut de Mathématiques et de Sciences Physiques (IMSP) at Dangbo, Benin, who has put in place a light organisation to lead the whole process. Two supervisors directed the experimentation as trainers. Even though the two supervisors were daily in contact with the sellers, five IMSP students volunteered to act as a liaison with the sellers and the broker using PGSs. The students handed the sellers' Android devices to right person every morning, stayed with them whenever possible and collected the devices back at the end of each day. This experimentation did not include interfacing mobile payment system; rather, PGS XML register was used as a reference to physically transfer cash when validated purchases between two parties reached the threshold to trigger a money transfer. However, local GSM operators and banks (MTN, MOOV, and BGFIBANK) clearly acknowledged the need to enter a process that will improve digital inclusion, and have then accepted to support the experimentation phase.

## 9. Conclusions and Future Work

The continuing growth of mobile devices in developing countries represents a tremendous opportunity for economic development. The features of this new potential market differ completely from the ones of the Western world, where, for example, companies in Silicon Valley are competing to develop the "killer app" in a certain category. In developing countries, innovators need to take the design of their applications from a different viewpoint, considering different cultural contexts and the motivations, experiences, needs of end-users. With PGS, we made the case for centring new systems on *trust models* already in place, and accompany these models' spontaneous evolution. The system we proposed does not require any specific technological skill for its use, and relies on powerful, easy to understand metaphors. At the same time, PGS relies on proven security techniques such as visual cryptography. Our future work focuses on moving from experimentation to large scale pilot deployment of PGS in a number of locations in West and East Africa.

## Acknowledgements

The project has been partly funded by the Bill and Melinda Gates Foundation, Grant Number GCE Phase I, ID # APP198273.